\documentclass{article}
\usepackage{spconf,amsmath,graphicx}
\usepackage{xcolor}
\usepackage{bm}
\usepackage{amsfonts} 
\usepackage{enumitem}
\setlist{nosep, leftmargin=14pt}

\usepackage{mwe} 

\title{Maximum Mean Discrepancy Kernels for Predictive and Prognostic Modeling of Whole Slide Images }
\name{Piotr Keller*, Muhammad Dawood* and Fayyaz ul Amir Afsar Minhas \thanks{* Joint first authorship}}

\address{Tissue Image Analytics Center, University of Warwick, United Kingdom}
\begin{document}
%
\maketitle
\begin{abstract}
How similar are two images? In computational pathology, where Whole Slide Images (WSIs) of digitally scanned tissue samples from patients can be multi-gigapixels in size, determination of degree of similarity between two WSIs is a challenging task with a number of practical applications. In this work, we explore a novel strategy based on kernelized Maximum Mean Discrepancy (MMD) analysis for determination of pairwise similarity between WSIs. The proposed approach works by calculating MMD between two WSIs using kernels over deep features of image patches. This allows representation of an entire dataset of WSIs as a kernel matrix for WSI level clustering, weakly-supervised prediction of TP-53 mutation status in breast cancer patients from their routine WSIs as well as survival analysis with state of the art prediction performance. We believe that this work will open up further avenues for application of WSI-level kernels for predictive and prognostic tasks in computational pathology.
\end{abstract}
\section{Introduction}
\label{sec:intro}

%
Computational Pathology (CPath) involves predictive modeling of routine histology Whole Slide Images (WSIs) for various diagnostic and prognostic tasks such as detection of cancer, discovery of image patterns associated with various genetic mutations and survival analysis which are conventionally, performed using deep learning approaches~\cite{huiQuPaper,atentionMIL}. Due to the sheer size of multi-gigapixel WSIs and the associated memory requirements of such predictive problems, WSIs are divided into patches and passed through a neural network to generate patch level prediction scores. These scores are then aggregated to generate whole slide level predictions. The issue with this widely used strategy is that it is inherently unable to capture information distributed over scales greater than the patch size. If we want to capture higher level structural information, we need to analyze larger regions or perform different types of aggregation ranging from simple averaging or max-pooling~\cite{maxPooling} to more sophisticated hierarchical attention models~\cite{HAG} and graph neural networks~\cite{lu2022slidegraph+}. 

\begin{figure*}[htb]
  \begin{center}
      \includegraphics[width=0.8\textwidth, ]{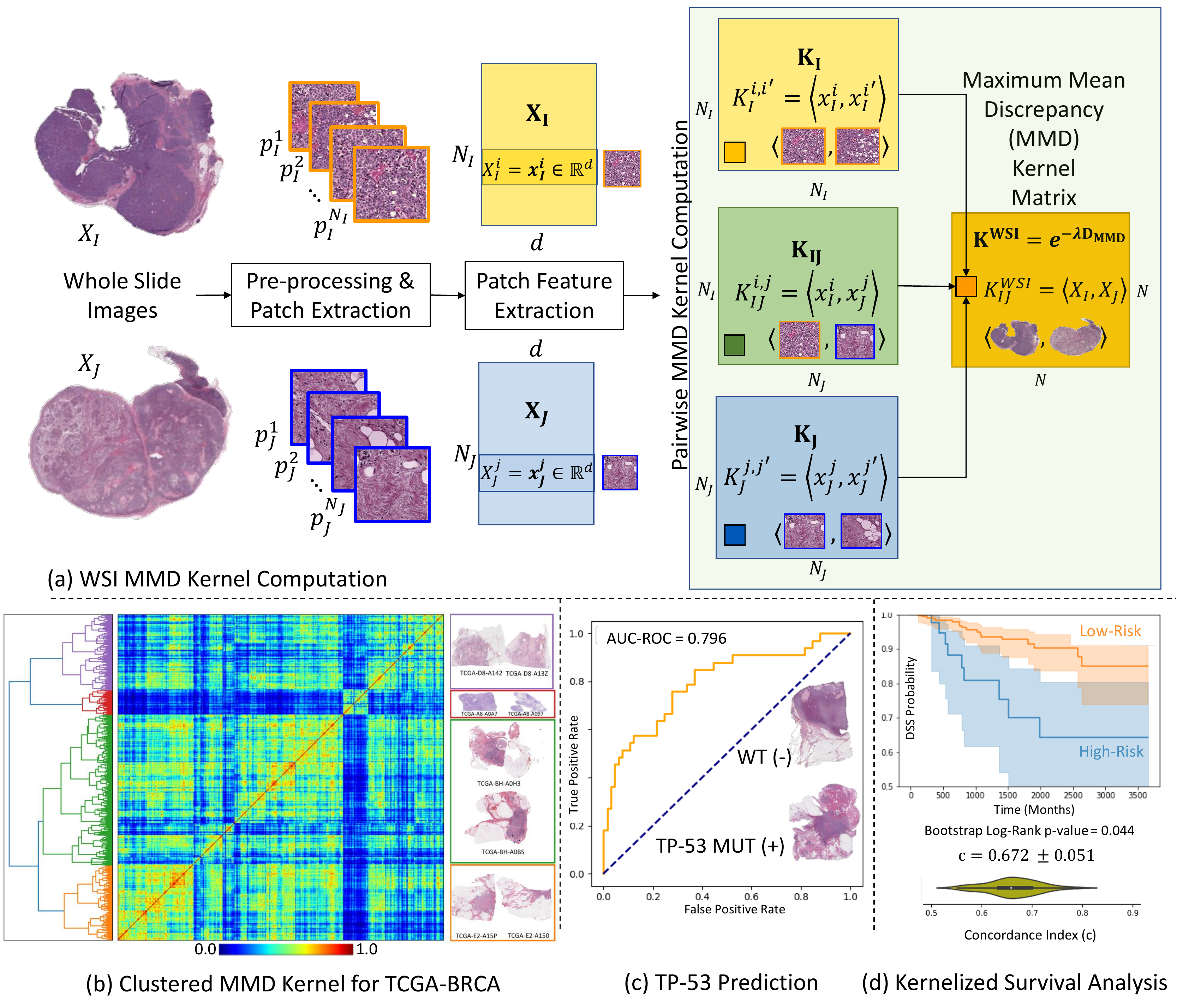}
        \hfill
        \caption{Proposed workflow and key results. (a) WSI data is pre-processed and non-overlapping image patches are obtained for patch-level  1024-dimensional feature extraction from a pre-trained ShuffleNet. The novel WSI-level Maximum Mean Discrepancy matrix and the corresponding MMD Kernel is computed (Eq. \ref{eq:MMDNew}). (b) The resulting MMD kernel for TCGA-BRCA dataset  with hierarchical clustering to visualize WSI-level groupings within the dataset. State of the art prediction results for TP-53 mutation status prediction (in (c)) and survival analysis (in (d)) using the proposed MMD kernel over TCGA-BRCA. }
        \label{fig:res}
    \end{center}
\end{figure*}

In this work, we approach the aggregation problem from a different perspective - the use of Maximum Mean Discrepancy (MMD) as a measure of (dis)similarity between WSIs (see Fig.~\ref{fig:res}). MMD is rooted in kernel matrix theory and is used as a statistical test to determine if two sets of measurements are drawn from different distributions or not~\cite{kernelTwoSampleTest}. The primary hypothesis underlying this work is that (dis)similairity between two WSIs can be captured by their MMD by simply representing each WSI as a set of features of its constituent patches. This not only enables computation of pairwise WSI (dis)similarity, the resulting metric also leads to formulation of WSI-level kernels thus opening up the possibility for the use of the existing repertoire of  powerful and versatile kernel methods for predictive modeling. Such kernels can capture slide level similarity in a dataset of WSIs and thus can be used for clustering, classification or survival prediction as well as integration of multi-modal data. To the best of our knowledge, this is the first report of WSI level kernels in this domain and it paves the way for using such methodologies for other related tasks.

Our primary technical contribution in this work is the proposed use of WSI-level MMD kernels for solving both unsupervised classification problems such as clustering and data mining as well as weakly-supervised predictive and prognostic tasks in CPath in which training label or target information is only available at the WSI-level. As a proof of principle, we use this approach to solve two clinically relevant problems - prediction of TP-53 mutation status and survival analysis of breast cancer patients using their routine Hematoxylin and Eosin (H\&E) stained tissue slides~\cite{breastCancerDeaths}. Below, we introduce both of the these problems along with a brief description of existing work in this domain. 

In the context of breast cancer diagnosis, detection of mutations in tumour suppressor genes such as TP-53 is of prime importance~\cite{p53Role}. TP-53 mutations, found in 20–40\% of all breast cancer cases, can be determinants of underlying cancer pathways and treatment options~\cite{tp53Percentage, best_treatment}. Previous work has shown that while it is possible to predict TP-53 mutation status from WSIs, the accuracy of such predictions is quite low with state of the art area under the receiver operating characteristic curve (AUC-ROC) of around 0.72~\cite{huiQuPaper}. 

Another important yet challenging problem in CPath is survival analysis which aims to analyze the expected duration of time until an event such as progression or death occurs. ~\cite{survivalPredClinitians} Current state of the art models for survival prediction from routine WSI data achieve a relatively low Concordance Index (C-Index) of no more than 0.58~\cite{mcat,atentionMIL,deepAtentionMIL}. 

\section{Materials and Method}
\label{sec:method}

\subsection{Study dataset}
We collected 1,133 WSIs of Formalin-Fixed paraffin-Embedded (FFPE) Hematoxylin and Eosin (H\&E) stained tissue section of 1084 breast cancer patients from The Cancer Genome Atlas~\cite{hoadley2018cell,Koboldt2012} (TCGA-BRCA). Inline with the previous study with state of the art results for point mutation prediction~\cite{huiQuPaper}, we filtered out WSIs 1) with poor visual quality; (2)  abnormal staining 3) with low informative tissue region; and (4) missing baseline resolution information. After slide selection, we end-up with 652 WSIs belonging to 652 patients. The TP-53 mutation status data for these patients was downloaded from CBioPortal (417 negative, 235 positive)~\cite{cerami2012cbio,gao2013integrative}. 

For survival analysis, we used the same dataset. However, previous state of the art approaches in this domain had used the whole dataset for analysis. In order to have a meaningful comparison, we have used the same cohort of patients. The Disease Specific Survival (DSS) data for these patients were collected from the TCGA Pan-Cancer Clinical Data Resource (TCGA-CDR)~\cite{liu2018integrated}. Patients with missing DSS data was not used in the study. In total we used WSI of 1,033 patients in our study. 

\subsection{Pre-processing and Representation of WSIs}
 Let $D = \{X_1, X_2,...,X_N\}$ be the set of $N$ WSIs in the dataset. We segment the tissue region of a WSIs using a tissue segmentation model to remove artefacts such as pen markings, tissue folding, etc. As an entire WSI at full resolution can be very large in size ($100,000 \times 150,000$ pixels) and does not fit into a GPU memory, we tile each WSI into patches of size $ 512 \times 512$ at a spatial resolution of 0.50 microns-per-pixel (MPP). Patches capturing less that 40\% of informative tissue area (mean pixel intensity above 200) are discarded, and the rest of patches (both tumour, and non-tumour) are used after stain normalization. We model a given WSI (indexed by $I$) as a set of patches as $X_I = \{x^{1}_{I}, x^{2}_{I},...,x^{N_I}_{I} \}$ such that $N_I$ is the number of patches in the image and $x^{i}_{I}$ represents the $i$th patch in image $X_I$. For each patch, we obtain its $d=1024$-dimensional feature representation $\bm x_i^I \in R^d$ by passing it through a convolutional neural network (CNN) encoder. Any feature extractor can be used as patch level encoder. However for this work, we extracted the latent representation of the second-last fully connected layer of an ImageNet pretrained SuffleNet~\cite{shuffleNet}.

\subsection{Maximum Mean Discrepancy Kernels between WSIs}
Modelling WSIs as sets allows us to use distance metrics defined over sets such as MMD to calculate the distance between any two WSIs. Originally proposed as a kernel two-sample test~\cite{kernelTwoSampleTest},  MMD between two WSIs can be defined as~\cite{geomloss}:
\begin{equation}\label{eq:MMD}
MMD(X_I,X_J) = \| \mathbb{E}_{x^i_I \sim X_I}[\varphi(x^i_I)] - \mathbb{E}_{x^j_J \sim X_J}[\varphi(x^j_J)]\|_\mathcal{H}
\end{equation}

Here, we assume that the WSIs $X_I$ and $X_J$ come from some space $\mathcal{X}$. We also have an abstract function $\varphi ; \mathcal{X} \mapsto \mathcal{H}$ and $\mathcal{H}$ is the RKHS that allows us to perform the change of basis on the patches. Thus, MMD measures distance in various moments of feature distributions of the patches in the two WSIs. Specifically, defining $\varphi(x)=x$ will result in the MMD between the two WSIs being simply the distance between average patch representations of the two WSIs: 
$MMD(X_I,X_J) = \| \mathbb{E}_{x^i_I \sim X_I}[x^i_I] - \mathbb{E}_{x^j_J \sim X_J}[x^j_J]\|_{R^d}=\|\mu_I-\mu_J\|$. Similarly, defining $\varphi(x)=(x,x^2)$ enables the resulting MMD to capture differences in means (as with $\varphi(x)=x$) as well as variances (second order moments) of corresponding features. It is important to emphasize that even though the approach is called Maximum "Mean" Discrepancy, through the use of kernels, it can calculate arbitrary higher order differences as discussed below. 

By varying our definition of $\varphi$ we can compare an arbitrary WSI-level statistical moment. However, it is often the case that two different distributions share some moments and thus may be seen as identical by MMD when in reality they are not. To combat this we would ideally like to compare an infinite number of moments which can be done by using kernels as these allow for implicit calculations of moments. To achieve this, we use the patch-level Guassian kernel defined as: $ k(x^i_I,x^j_J) = e^{\frac{-\| x^i_I-x^j_J \|^2}{4 \sigma^2}}$ where $\sigma$ is the standard deviation (also called blur parameter) of the kernel which defines how large the region of influence of a particular point is, the larger the $\sigma$ the bigger the region. The benefit of using kernels comes from the Moore–Aronszajn theorem~\cite{mooreAronszajnTheorem} which states that a kernel such as the one defined above has an implicit feature representation given by:
\begin{equation}
k(x^i_I,x^j_J) = \langle x^i_I, x^j_J\rangle = \varphi(x^i_I)^T\varphi(x^j_J)
\end{equation}

As it is possible to rearrange equation \ref{eq:MMD} to be only expressed in terms of dot products, we can simply use $k(x^i_I,x^j_J)$ instead of calculating the actual dot product after the change of basis to $\mathcal{H}$. The modified MMD equation that makes use of kernels can be defined as:

\begin{equation} \label{eq:MMDNew}
\begin{split}
MMD^2(X_I,X_J) & = \mathbb{E}_{X_I}[k(x^i_I,x^{i'}_I)] + \mathbb{E}_{X_J}[k(x^j_J,x^{j'}_J)] \\
 & - 2 \mathbb{E}_{X_i,X_j}[k(x^i_I,x^j_J)]
\end{split}
\end{equation}

With this, we have now defined a way of calculating pairwise distance between any two WSIs with the same patch level representation.

With the ability to calculate pairwise distances between two WSIs, we can generate an $N\times N$ distance matrix, $\bm D^{MMD}$, which stores all distances between all pairwise combinations of WSIs such that $\bm D^{MMD}_{I,J}=MMD^2(X_I,X_J)$ for $I,J = 1 \ldots N$. This means the relationship of the entire dataset can be represented by a single matrix that can fit into memory. We can get a corresponding symmetric and positive semi-definite Mercer kernel matrix from this distance matrix as follows: 

\begin{equation}
    \bm K_{\gamma}=e^{-\gamma D^{MMD}}
\end{equation}

where $K_{I,J} = \langle X_I, X_J \rangle = \psi(X_I)^T\psi(X_J)  \in [0,1]$ is the similarity between $X_I$ and $X_J$ under an implicit feature representation $\psi$ and $\gamma \in \mathbb R_{\ge 0}$ is a model hyper-parameter.

\subsection{Implementation \& Applications of MMD Kernels}
To generate the MMD kernel we used GeomLoss as it provided us with an efficient GPU implementation to calculate MMD~\cite{geomloss}. The complete implementation is available at:
\color{blue}{https://github.com/engrodawood/Hist-MMD}. \color{black} We can use this WSI level kernel for any predictive or analytical task. The development of WSI kernels enables using the vast and powerful repertoire of kernel methods in computational pathology ranging from agglomerative clustering, support vector clustering~\cite{supportVectorClustering},  kernel principal component analysis, support vector regression, classification and survival analysis. In addition it opens up the possibility of using multiple kernel learning and cross-domain feature integration. As discussed below, we demonstrate the effectiveness of the proposed approach for TP-53 mutation status prediction and survival analysis.  

\subsection{TP-53 Mutation prediction and evaluation}
\label{sec:TP-53 predicton}
We assessed the significance of the proposed MMD kernels in predicting the mutation status of TP-53. To do this we used a binary class Support Vector Machine (SVM) with its precomputed kernel set to the MMD kernel defined above~\cite{scikit-learn}. 

For performance evaluation we used the same methodology as our deep-learning baseline method~\cite{huiQuPaper} which randomly partitioned the filtered TCGA slides into training, validation (used for model selection) and test sets based on 70\%, 15\%, and 15\% ratios, respectively. The models performance was measured by using AUC-ROC. The 95\% Confidence Interval (CI) of each AUC-ROC score is calculated by 1000 bootstrap runs to estimate the uncertainty of the result. The hyperparamters chosen for the kernel (through validation) are $\gamma=4, \sigma=10$ whereas the cost control parameter of the SVM was set as $C=10,000$.

\subsection{Kernalized Survival Analysis}
We have used an existing implementation of kernel survival SVMs (KSSVM)~\cite{fastTrainingSurvivalAnalysis} to demonstrate the effectiveness of the proposed approach for survival analysis. KSSVM predicts a risk score  $f(X)$ for a given test patient using their WSI representation $X$ after being trained over a training dataset $\{(X_I,T_I,\delta_I)|i=1\ldots N_{train}\}$ which consists of tuples of a patient's whole slide image $X_I$, their disease-specific survival time $T_i$ and a binary event indicator variable $\delta_I \in \{0,1\}$ which shows whether the patient has passed away from breast cancer within a censoring time $T_{censor} = 10$ years or not. The implementation of KSSVM allows use of an implicit or pre-computed kernel and has a single hyperparameter $\alpha$ which controls the loss penalty term in its objective function (see~\cite{fastTrainingSurvivalAnalysis} for further details). We chose, $\alpha=0.125$ and $\gamma$ is set to simply the multiplicative inverse of the median of $D^{MMD}$ to reduce the number of tunable hyper-parameters as the effect of $\gamma$ is also modulated by the blur parameter of the patch level kernel. 

For performance evaluation we used out-of-bag bootstrapping to generate the training and test datasets R = 50 times stratified with respect to the survival event. In each run we generate the risk scores of the test examples which are then used to compute C-index as our primary performance metric. C-index measures the degree of concordance between relative prediction scores of test patients and their actual survival times. In line with AUC-ROC, the C-index ranges from 0.0 (inverted ranking of survival scores) to 0.5  (no concordance between predicted scores and actual survival times) to 1.0 (perfect concordance between prediction scores and actual survival times). For a representative bootstrap run, we show the Kaplan-Meier survival curves of the high risk and low risk patient groups using an optimal threshold computed from the training set scores along with p-value of the log-rank test across all bootstrap runs as: $2\times median(\{p_r|r = 1\ldots R\})$. 

\section{Results}

\begin{table}[htb]
\begin{center}
\begin{tabular}{ c|c } 
 \hline 
 Method & Concordance Index $\pm$ std dev \\ 
 \hline & \\[-2.5ex]
 MMD Kernel (Proposed) & 0.672 $\pm$ 0.051\\
 MCAT~\cite{mcat} & 0.580 $\pm$ 0.069\\ 
 Attention MIL~\cite{atentionMIL} & 0.564 $\pm$ 0.050\\ 
 DeepAttnMISL~\cite{deepAtentionMIL} & 0.524 $\pm$ 0.043\\ 
\end{tabular}
\caption{\label{tab:survivalTable} Comparison of C-index $\pm$ std dev.}
\end{center}
\end{table}

The key results using the proposed approach are given in Fig.~\ref{fig:res} with the entire TCGA-BRCA kernel matrix and the resulting hierarchical clustering showing clear similarities between WSIs in the same cluster. 

For TP-53 point mutation prediction, we obtained an AUC-ROC (with 95\% CI) of 0.7958 (0.6966-0.8863) in contrast to 0.7288 (0.6205-0.8278) by the baseline study. 

For survival prediction we can see in Table \ref{tab:survivalTable} that we achieved a much higher C-Index $\pm$ standard deviation of $0.67 \pm 0.05$ in contrast to the current state of the art of $0.580 \pm 0.069$. The proposed approach also allows risk stratification with Kaplan-Meier analaysis with a statistically significant bootstrap p-value of the log-rank test at $0.044$. 

The results show the promise of the proposed kernel  as we have achieved a significantly better performance then the state of the art for two clinically important problems. 


\section{Conclusions and Future work}
Beyond the application to the specific machine learning problems discussed above, the proposed MMD WSI-kernels can be used for other predictive and prognostic modelling tasks. We believe that this work will open the avenue for further research on the use of previously under-explored yet powerful kernel methods for WSI in computational pathology. In the future we will investigate the usefulness of this method on independent datasets. Secondly, while the use of kernel methods provides WSI-level similarity metrics, further approaches for interpretability and explainability of these results need to be investigated. In the future, we will also investigate the integration of multi-modal and cross-domain kernels. 

\section{Compliance with ethical standards}
\label{sec:ethics}

This research was conducted retrospectively using pen access human subject data by TCGA~\cite{hoadley2018cell,Koboldt2012}. Ethical approval was not required by TCGA/CBioPortal license.

\section{Acknowledgments}
\label{sec:acknowledgments}
FM acknowledges funding from EPSRC EP/W02909X/1 and PathLAKE consortium. MD and FM report research funding from GlaxoSmithKline outside the submitted work.

\bibliographystyle{IEEEbib}
\bibliography{strings,refs}

\end{document}